# Spin-Polarized Tunneling through Chemical Vapor Deposited Multilayer Molybdenum Disulfide

Andre Dankert[1†], Parham Pashaei[1], M. Venkata Kamalakar[1,2], Anand P.S. Gaur[3], Satyaprakash Sahoo[3,4], Ivan Rungger[5,6], Awadhesh Narayan[6,7], Kapildeb Dolui[5,8], Anamul Hoque[1], Michel P. de Jong[9], Ram S. Katiyar[3], Stefano Sanvito[6], Saroj P. Dash[1]*

[1] *Department of Microtechnology and Nanoscience, Chalmers University of Technology, SE-41296, Göteborg, Sweden.*
[2] *Department of Physics and Astronomy, Uppsala University, Box 516, 75120, Uppsala, Sweden*
[3] *Department of Physics and Institute for Functional Nanomaterials, University of Puerto Rico, San Juan, PR 00931, USA.*
[4] *Institute of Physics, Bhubaneswar, Odisha 751005, India.*
[5] *School of Physics, AMBER and CRANN Institute, Trinity College, Dublin 2, Ireland.*
[6] *National Physical Laboratory, Teddington, TW11 0LW, United Kingdom*
[7] *Department of Physics, University of Illinois at Urbana-Champaign, Urbana, Illinois 61801, USA*
[8] *Department of Physics and Astronomy, University of Delaware, Newark, Delaware 19716-2570, USA*
[9] *MESA+ Institute for Nanotechnology University of Twente 7500 AE Enschede, the Netherlands*

Email: †andre.dankert@chalmers.se; *saroj.dash@chalmers.se

**Abstract**

The two-dimensional (2D) semiconductor molybdenum disulfide ($MoS_2$) has attracted widespread attention for its extraordinary electrical, optical, spin and valley related properties. Here, we report on spin polarized tunneling through chemical vapor deposited (CVD) multilayer $MoS_2$ (~7 nm) at room temperature in a vertically fabricated spin-valve device. A tunnel magnetoresistance (TMR) of 0.5 – 2 % has been observed, corresponding to spin polarization of 5 - 10 % in the measured temperature range of 300 – 75 K. First principles calculations for ideal junctions results in a tunnel magnetoresistance up to 8 %, and a spin polarization of 26 %. The detailed measurements at different temperatures and bias voltages, and density functional theory calculations provide information about spin transport mechanisms in vertical multilayer $MoS_2$ spin-valve devices. These findings form a platform for exploring spin functionalities in 2D semiconductors and understanding the basic phenomenon that control their performance.

Spintronics is an emerging field for "beyond-CMOS" technology, where information is carried by spin instead of charge.[1] One of the primary challenges in this field is to discover novel semiconductor materials with functional spintronics properties.[2,3,4] Two dimensional (2D) crystals offer a unique potential for spintronic devices due to remarkable properties such as long spin-coherence lengths,[5,6] spin-polarized tunneling,[7,8] high spin-orbit coupling (SOC)[9] and spin-momentum locking.[10,11] Recently, long spin coherence length in graphene has been achieved, due to its low SOC strengths and high mobility.[6,12,13] At the same time, atomically thin molybdenum disulfide ($MoS_2$) has emerged as a promising semiconducting 2D crystal, demonstrating novel electronic, optoelectronic and spintronic properties. In monolayer $MoS_2$, the lack of inversion symmetry coupled to the high spin-orbit interaction leads to a unique spin and valley polarization.[14] Recently, nanosecond electron spin lifetimes have been observed in monolayer $MoS_2$ at low temperatures, by using optical Kerr spectroscopy experiments.[15] However, electrical realization of lateral spin transport in a $MoS_2$ channel remain challenging.[16]

The investigation of spin transport in a vertical magnetic tunnel junction (MTJ) is an interesting approach, where the transport channel is defined by a few nanometer-thick $MoS_2$ spacer sandwiched between two ferromagnetic (FM) electrodes. Employing such MTJs, a unique spin filtering effect was theoretically predicted for some 2D materials and first observed in graphene/graphite based devices.[17,18,19,20] Beyond graphene, it is interesting to investigate whether similar spin filtering effects may be observed for MTJs based on insulating or semiconducting 2D crystals. Recently, insulating hexagonal boron nitride (h-BN) tunnel barriers have been used showing excellent spin polarized tunneling and filtering properties.[21,22,7,8] Since h-BN binds little to the metallic surfaces, one may wish to fabricate MTJs with other layered compounds likely to be more reactive with the magnetic electrodes. This is for instance the case of $MoS_2$, since it was predicted that semiconducting multilayer $MoS_2$ junctions can exhibit a large TMR up to 300% in the tunneling regime.[23] Recently, *ab initio* calculations and experiments have shown that monolayer $MoS_2$ and $WS_2$ MTJs are metallic due to strong coupling between the Fe and the S atoms at the interface, showing a magnetoresistance of ~ 0.5 %.[24,25,26]. However, MTJs incorporating multilayer $MoS_2$ with semiconducting properties are expected to show enhanced performance. Such multilayered $MoS_2$ MTJs and room temperature operation have not been realized experimentally so far. Moreover, the possibility of using large scale multilayer $MoS_2$ can further enhance the impact of such devices for practical applications.

Considering the potential impact of 2D semiconductor MTJs, here we report on spin polarized tunneling through a CVD grown multilayer $MoS_2$ in a spin-valve structure, measuring a tunnel magnetoresistance (TMR) up to room temperature. Specifically, we have employed vertical spin-valve devices, with 7 nm thick multilayer $MoS_2$ sandwiched between two FM electrodes. A TMR up to 2 % is observed when the magnetic alignment of the two electrodes is switched from parallel to anti-parallel due to the transmission of spin polarized electrons through the multilayer MoS$_2$ spacer. By combining bias and temperature dependent TMR measurements with density functional theory calculations, we bring out the detailed information about the spin polarization at the interfaces and the spin transport process through the multilayer MoS$_2$ junctions.

**RESULTS AND DISCUSSION**

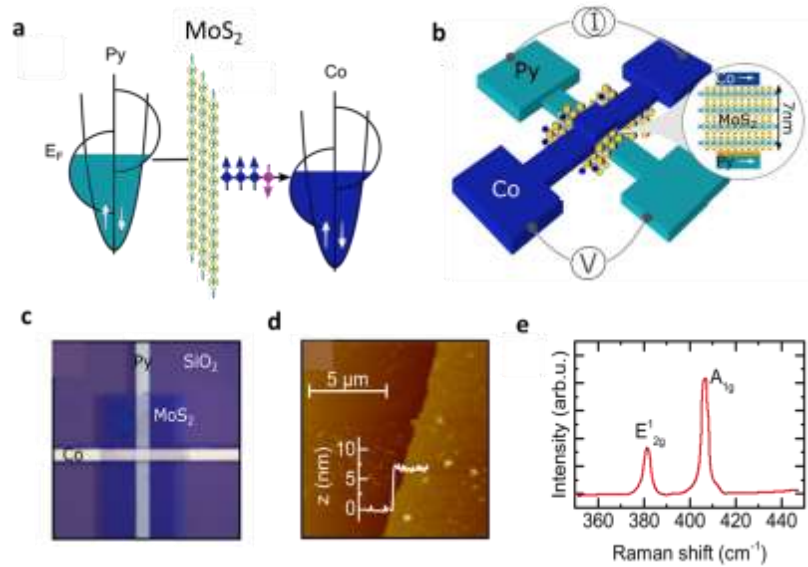

*Figure 1*. Multilayer MoS$_2$ tunnel magnetoresistance device. *a,* Spin-dependent tunneling in magnetic tunnel junctions with multilayer MoS$_2$ barrier. *b,* Schematic representation of the multilayer MoS$_2$ vertical device with ferromagnetic contacts and MoS$_2$ spacer. The measurement scheme is shown with four-probe cross bar geometry. *c,* Optical microscope image of a fabricated device consisting of large area CVD grown multilayer MoS$_2$ junction of 7 nm thickness and ferromagnetic Co and Ni$_{80}$Fe$_{20}$ (Py)/AlO$_x$ (0.8 nm) contacts as top and bottom electrodes respectively. The active junction area is 5× 20 μm². *d,* Atomic force microscope scan of 7 nm CVD MoS$_2$ on SiO$_2$/Si substrate. *e,* Raman spectra of 7 nm CVD MoS$_2$ measured at room temperature.

Tunneling magnetoresistance (TMR) is a consequence of the spin-dependent tunneling in magnetic tunnel junctions with MoS$_2$ barrier as shown in Fig. 1a. The device geometry shown in Fig 1b incorporates a 7nm-thick MoS$_2$ layer sandwiched between two FMs in a vertical structure. Specifically, the devices consist of a Ni$_{80}$Fe$_{20}$ (30 nm)/AlO$_x$(0.8nm)/MoS$_2$(7nm)/Co(50 nm) stack, fabricated using photo-lithography, metal evaporation and 2D layer transfer techniques (see Methods). The thin AlO$_x$ layer was prepared by Al evaporation and a natural oxidation. This AlO$_x$ layer is expected to protect the bottom Ni$_{80}$Fe$_{20}$ electrode from oxidation and acts as a leaky tunnel barrier.[27] An optical microscope image of a fabricated MoS$_2$ vertical device and the measurement scheme is shown in Fig 1 c. The multi-layer MoS$_2$ chosen in our devices is grown over a large area on a SiO$_2$/Si substrate by CVD.[28] The thickness of the MoS$_2$ layer is determined to be 7 nm by AFM measurement as shown in Fig. 1d, which corresponds to around 10 monolayers (1 monolayer around 6.5 Å). Figure 1e shows the Raman spectrum of a MoS$_2$ film displaying the two Raman active modes at ~384 cm$^{-1}$($E^1_{2g}$) and ~407 cm$^{-1}$ (A$_{1g}$) at room temperature.

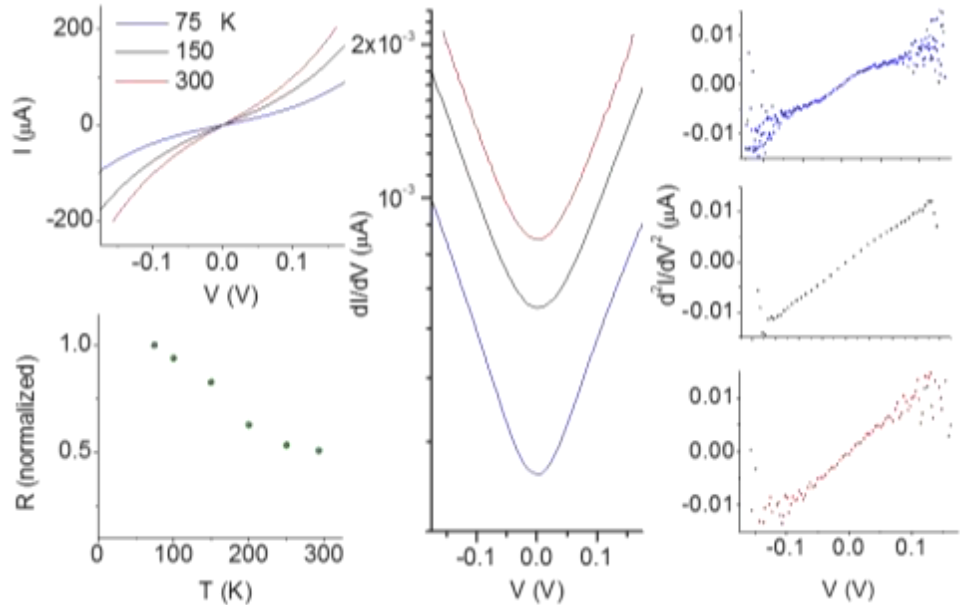

**Figure 2.** *Electrical characterization of multilayer MoS$_2$ vertical devices. **a**, Current-voltage (J-V) characteristics of the junction at different temperatures. **b**, Temperature dependence of MoS$_2$ junction resistance (normalized) at bias voltage of 5 mV. **c,** The resistance versus bias voltage curve of the junction at temperatures of 75 K and 300 K.*

The electrical transport properties of our junctions are measured in a four-terminal geometry as displayed in Fig. 1b. We observe increasingly nonlinear current-voltage (*I-V*) characteristics of the junction at lower temperatures (Fig. 2a). Figure 2b displays the normalized junction resistance (R=V/I) as a function of temperature (*T*), where the R value doubles when cooling down from 300 to 75 K. Such a large variation is expected due to the presence of a MoS$_2$ semiconducting barrier in the junction, whereas insulating Al$_2$O$_3$ barriers usually show an increase in R of 10-20% in the same range.[27] The strong temperature dependence can be attributed to inelastic tunneling or gap-state assisted tunneling through the MoS$_2$ layers.[29] The resistance versus bias voltage curves at both high and low temperature are quasi-symmetric as shown in Fig. 2c, signifying a deviation from a rectangular potential barrier. Considering different barrier heights for AlO$_x$ and MoS$_2$, it is reasonable to assume that the overall barrier is asymmetric. This is important for junctions made of multilayer $MoS_2$, which has a small band gap of only ~ 1.2 eV. The junctions are also found to be quite stable up to an applied bias of 50 mV and also do not exhibit any zero bias anomaly, suggesting the absence of magnetic impurities.

The junction resistance for thick $MoS_2$ (7 nm) consists of both the inter-layer and intra-layer resistances. The couplings between the $MoS_2$ layers are expected to arise from the overlap of the electron wave functions due to the small separation between the sulfide layers, with charge screening length of ~7 nm.[30] This is distinctly different from the current distribution in multilayer graphene with charge screening length of only 0.6 nm.[31] The difference arises due to the transport involving the d-electrons of $MoS_2$, while the $p_z$-orbitals are responsible in graphene. For thinner $MoS_2$ layers (few monolayers) a direct tunneling dominates, with the tunneling conductance exponentially decreasing with increasing the $MoS_2$ thickness.[32,33,34]. However, for thicker samples used in the present study, inelastic tunneling or gap-state assisted tunneling through defects in the form of S vacancies cannot be ruled out.[29]

Next, we performed magnetoresistance measurements with 7 nm multilayer $MoS_2$ MTJs by applying a fixed bias current (constant current mode) while measuring the voltage drop as a function of the external in-plane magnetic field, *B*. The TMR across the $MoS_2$ junction is observed at room temperature as a difference in the resistances measured for the parallel, $R_p$, and antiparallel, $R_{ap}$, alignment of the magnetizations of the FM electrodes as shown in Fig. 3a. The well-defined resistance states $R_p$ and $R_{ap}$ are achieved by using different FM materials (Co and NiFe) and electrode widths on either side of the $MoS_2$ layer. The measured magneto voltage signal of ΔV= 15 µV corresponds to a $TMR = \frac{R_p - R_{ap}}{R_p} \times 100\%$ = 0.5 % at 300 K. We estimate the spin polarization of the contacts from the Julliere relation $TMR = \frac{2P_1 P_2}{1 - P_1 P_2}$,[35,36] and these turn out to be $P_1$ = $P_2$ = 5% (assuming the polarizations of Co and $Ni_{80}Fe_{20}$ to be the same). We would like to note that the spin polarization obtained with introduction of $MoS_2$ is smaller than the polarization of ferromagnetic tunnel contacts reported in literature.[36] This can be attributed to disorder introduced at the interfaces during the wet transfer process of the CVD $MoS_2$ layer onto the ferromagnetic electrode. It has also to be noted that the spin transport process is very sensitive to defect assisted tunneling (i.e. multi-step tunneling) through $MoS_2$ and can significantly affects the TMR.[37] This can cause additional spin flip scattering due to the presence of strong spin-orbit coupling in $MoS_2$.[38] As predicted theoretically, the spin polarization is expected to be higher with development of all in-situ method for preparation of good quality $MoS_2$ layers and ferromagnetic tunnel contacts. However, our effort to integrate ferromagnetic tunnel junction to CVD $MoS_2$ in an ex-situ fabrication process shows promising results with a TMR up to room temperature.

Even though the observation of a TMR offers evidence for spin-polarized transport, a careful investigation of the bias and temperature dependence is essential for physical understanding of the phenomenon. The applications of negative and positive bias voltages correspond to a spin transport through the $MoS_2$ from $Ni_{80}Fe_{20}$ to Co and *Co* to $Ni_{80}Fe_{20}$ respectively. The bias dependence of the TMR signal changes sign with reversing bias polarity (shown in Fig. 3a). The bias dependence of the absolute value of TMR at room temperature measured at different voltages across the junction is shown in Fig. 3b. We observe a decrease of the TMR at higher bias voltages with a maximum around zero bias voltage. Such behavior can be attributed to the excitation of magnons, to band bending and possibly to the involvement of interface states at high voltages, as also observed in other material systems.[39] The bias dependence is also found to be asymmetric with polarity, decreasing much faster in the negative than the positive bias range. This behavior can be due to asymmetric barrier interfaces ($Co/MoS_2$ and $Ni_{80}Fe_{20}/AlO_x/MoS_2$) at the two sides of the $MoS_2$ layer. Similar bias dependence behavior has also been observed in some cases in other magnetic tunnel junctions with inorganic and organic semiconductors,[40] and tunnel junctions with $Al_2O_3$, MgO and h-BN. [41,42,22]

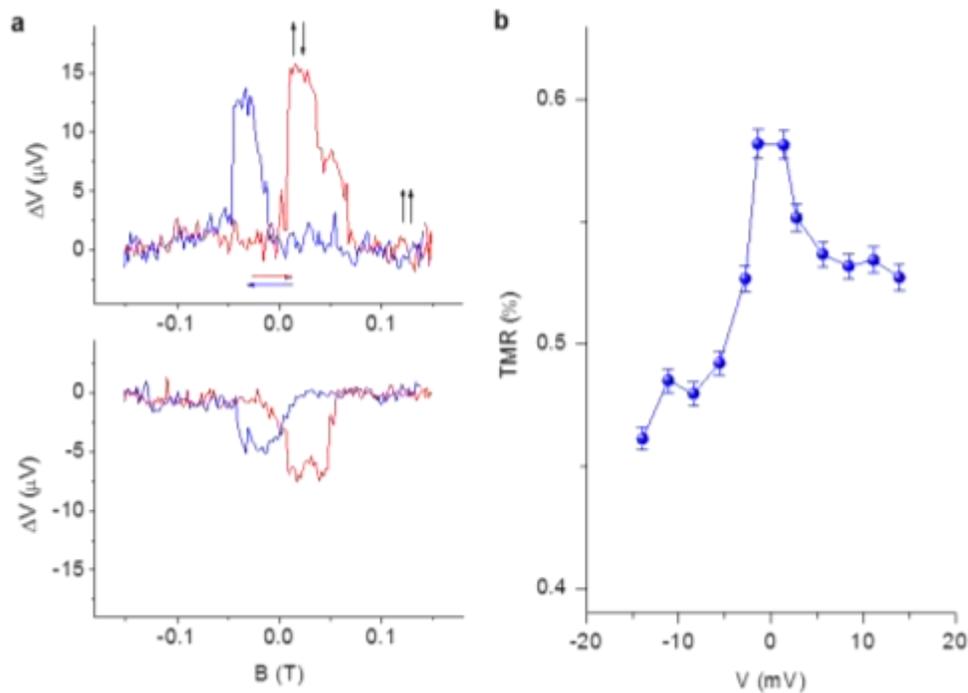

***Figure 3.*** *Spin-valve measurements in multilayer $MoS_2$ MTJs at room temperature.* ***a,*** *Tunnel magnetoresistance (TMR) measurements on the device with an in-plane magnetic field $B_{in}$ for applied bias voltages of +10 mV (upper panel) and -5 mV (lower panel) at room temperature (300 K). The arrows indicate the up and down B field sweep directions.* ***b,*** *Bias dependence of TMR signal measured at 300 K.*

In order to investigate the details of the spin transport process through $MoS_2$, spin-valve measurements were performed at different temperatures. The TMR shows an overall increase up to 2% at 75 K, corresponding to a spin polarization of 10% (as extracted from Julliere's formula[35]). Figure 4a shows the TMR at 75 K and 300 K and the normalized TMR as a function of temperature is plotted in Fig. 4b. We model the observed decrease in TMR with increase in temperature by considering the spin polarization to have the same temperature dependence as surface magnetization, which is described by the spin wave excitation model with $T^{3/2}$ dependence.[43] We observe a faster decrease of the TMR at higher temperatures, in comparison to the expected moderate decrease. Such behavior can be attributed to the low bandgap of ~ 1.2 eV and correspondingly low barrier height of the multilayer $MoS_2$ barrier.

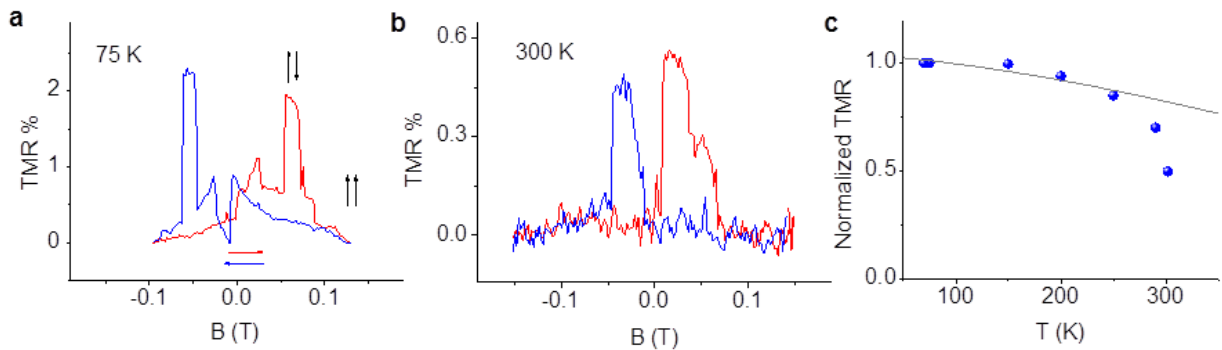

*Figure 4.* *Temperature dependence of TMR in multilayer $MoS_2$ MTJ.* ***a****, Tunnel magnetoresistance (TMR) measurements at 75 K.* ***b****. TMR measured at 300 K. The arrows indicate the up and down B field sweep directions.* ***c****, Temperature dependence of TMR signal (normalized). The fitting represents the temperature dependence of TMR α 1−α$T^{3/2}$.*

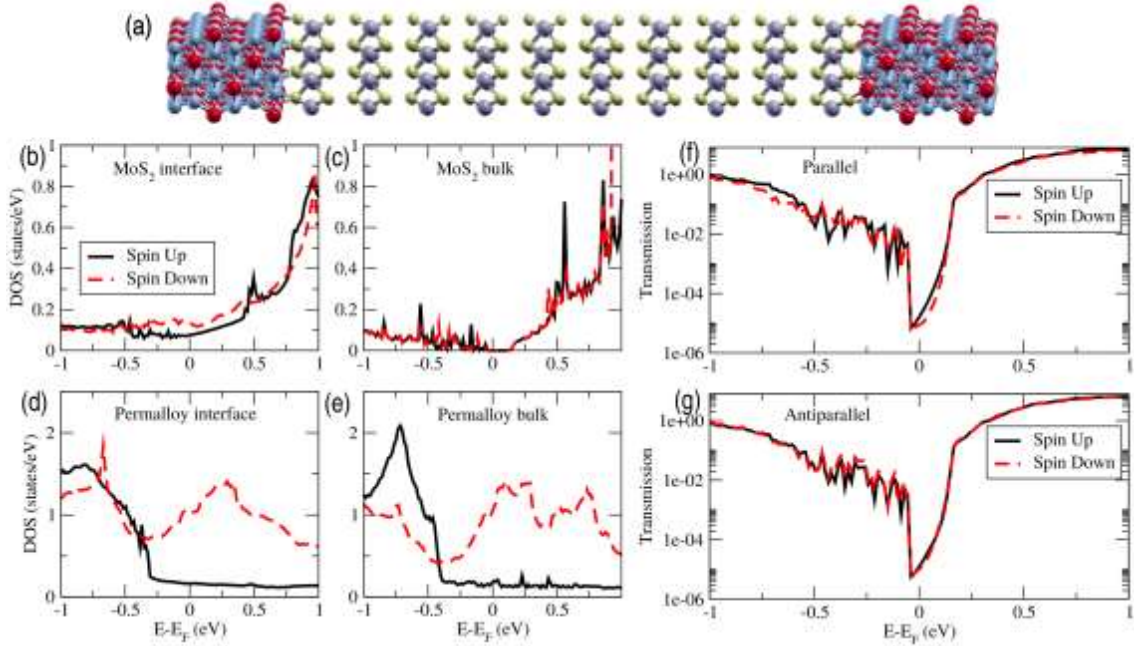

*Figure 5. Theoretical results for a permalloy/MoS$_2$/permalloy junction. (a) Junction setup for transport calculations with ten layers of MoS$_2$, corresponding to a barrier thickness of about 6.4 nm, sandwiched between semi-infinite permalloy (Ni$_{80}$Fe$_{20}$) electrodes. Density of states (DOS) as a function of energy for the (b) interface MoS$_2$ layer, (c) MoS$_2$ layer in the bulk of the spacer, (d) interface permalloy layer and (e) bulk permalloy. Spin resolved transmission in (f) for parallel and (g) antiparallel configurations of the junction.*

In order to further understand the transport properties of the devices, we have carried out density functional theory based transport calculations for an ideal, defect-free junction (see Computational Methods for details). The setup for our computations is shown in Fig. 5(a) for a junction made of ten MoS$_2$ layers sandwiched between two semi-infinite permalloy (Ni$_{80}$Fe$_{20}$) electrodes. Note that we use permalloy for both electrodes, which is a reasonable approximation since the Co density of states (DOS) is rather similar to the one of permalloy.[44] The use of an approximately symmetric setup is also justified by the rather small asymmetry found in the experimental *I-V* and TMR-*V* curves. The spin resolved DOS projected onto different layers of the junction are shown in Fig. 5(b)-(e). We find that the MoS$_2$ interface layer becomes metallic due to a strong hybridization with permalloy, similar to what has been found for single layer MoS$_2$.[24] In contrast, as one moves into the bulk of the spacer, a gap reminiscent of pristine MoS$_2$ emerges. Notably, the DOS at the Fermi level of the interface MoS$_2$ layer becomes spin polarized, which indicates spin injection into MoS$_2$. Considering the first MoS$_2$ layer as the effective metallic interface layer, we find the theoretical upper limit to the efficiency for spin injection in this junction, as η = (DOS$_↓$-DOS$_↑$) / (DOS$_↓$+DOS$_↑$) ~ 26%. This number is significantly smaller than the value in the bulk permalloy (73%) and at the permalloy interface layer (76%).

The spin resolved transmission as a function of energy for the parallel and antiparallel configurations is plotted in Fig. 5(f) and (g), respectively. At the Fermi level, we find MR = ($T_\text{parallel}$ - $T_\text{antiparallel}$)/ $T_\text{antiparallel}$ ~ 8%. This MR value, under ballistic transport conditions for an ideal tunnel junction, should be considered as an upper limit to the MR for devices with permalloy electrodes. The presence of defects in $MoS_2$, as well as inelastic scattering processes can reduce the spin polarization of the current, thereby reducing the measured magnetoresistance down to the experimentally measured values. Note that the MR obtained here is rather similar to that of a single $MoS_2$ layer with permalloy electrodes,[24] and significantly lower than what was predicted in a previous study for $MoS_2$ junctions with Fe electrodes,[23] where it was found to increase with $MoS_2$ thickness. The origin of the reduced MR found here lies in the reduced polarization and spin filtering induced by the permalloy electrodes when compared to Fe electrodes, which also have a better lattice match with $MoS_2$. This observation also shows that the choice of electrode materials and lattice matching are crucial factors in obtaining high magnetoresistance in $MoS_2$ based tunnel junctions.

**CONCLUSION**

In summary, we have demonstrated spin-polarized tunneling in multilayer $MoS_2$ at room temperature in a vertical spin-valve device. The spin-transport through 7 nm of $MoS_2$ produces a TMR of 0.5% at room temperature, which shows enhancement up to 2% at 75 K, corresponding to an increase of spin polarization from 5 to 10%. Our density functional theory based transport calculations for ideal spin valves provide an upper limit to the TMR to be 8%, while the maximum spin polarization is obtained to be 26%. The theoretical results also show that interfaces without epitaxial growth generally lead to rather low MR ratios when compared to junctions with better lattice match across the layers. The lower experimentally obtained spin polarization possibly can be attributed to interface contamination during transfer process, defect assisted tunneling and spin flip scattering due to spin-orbit coupling in $MoS_2$. These results on vertical spin-transport in large area multilayer $MoS_2$ reveal useful information needed for the development of 2D semiconductor materials and their heterostructures for spintronic devices and will open the path for the observation of novel spintronic effects. One particularly interesting new topic would be to investigate spintronic devices based on 2D material heterostructures by integrating the $MoS_2$ layers into graphene spin transport channels[4].

**EXPERIMENTAL METHODS**

**Materials growth and characterization.** We fabricated the $MoS_2$/ferromagnetic metal heterostructures using $MoS_2$ films grown by chemical vapor deposition (CVD). The large area $MoS_2$ films were synthesized on $SiO_2$/Si substrates via CVD and sulfurization of molybdenum films at 900 °C. The film thickness was characterized in tapping mode atomic force microscopy (AFM-VEECO). Raman and photoluminescence (PL) spectroscopy were carried out using Horiba-Jobin T64000 (triple mode subtractive) micro-Raman system in backscattering configuration utilizing Argon ion laser (514.5 nm line as excitation source).

**Device fabrication**. Spin transport devices incorporating a multilayer $MoS_2$ between two FM metal electrodes in a vertical geometry are fabricated. The bottom $Ni_{80}Fe_{20}$ (30 nm) electrodes are prepared on a Si/SiO$_2$ substrate by photolithography, electron beam evaporation and lift off methods. Before deposition of NiFe electrodes, we deposited a thin layer of Ti in order to increase adhesion of FM to the SiO$_2$ substrate. The bottom NiFe electrodes are capped with 0.8 nm Al and subsequently oxidized naturally to make an AlO$_x$ layer, which should protect the bottom ferromagnet NiFe from oxidation and contamination during transfer of MoS$_2$.

The large area MoS$_2$ films grown on SiO$_2$ substrate were first covered with PMMA layer and then released from the substrate by etching in KOH. After a de-ionized (DI) water rinse, the MoS$_2$/PMMA layer was transferred onto the bottom Ni$_{80}$Fe$_{20}$ electrodes. After drying the chip in an ambient environment, we annealed it at 150 °C for 10 min. It has been observed that this annealing step improved adhesion of MoS$_2$ with the bottom Ni$_{80}$Fe$_{20}$ electrode. The chip was then cleaned with acetone to remove the PMMA. The active device areas of MoS$_2$ layer were patterned by lithography and Ar ion beam etching with SIMS etch stop technique. The top Co (65 nm) electrodes and capping Au (20 nm) layer were prepared by photolithography, electron beam evaporation and lift off techniques in a cross-bar geometry. The final device consists of junctions with Ni$_{80}$Fe$_{20}$ (30 nm)/AlO$_x$ (0.8 nm)/MoS$_2$ (7 nm)/Co (65 nm) heterostructures.

**Computational Methods.** Ab initio transport calculations were performed using the *Smeagol* package,[45,46,47]. The code interfaces non-equilibrium Green's function method for transport with density functional theory, as implemented in SIESTA code.[48] Norm conserving pseudopotentials were used to replace the core electrons and a double ζ polarized basis set was employed, along with a mesh cutoff of 300 Rydberg. Local density approximation to the exchange correlation functional was used.[49] An in-plane supercell of dimensions 14.37 Å x 8.29 Å was constructed, with the total length of the scattering region along the transport direction being 82.65 Å. Periodic boundary conditions were implemented perpendicular to the transport direction, with a 2x4 *k*-point mesh sampling used for the self-consistent calculations. Transmission coefficients and densities of states were obtained for the so converged charge density by integrating over a denser 10x20 *k*-point grid.

## Acknowledgement

S. Dash acknowledges financial support from the NanoAoA, Swedish Research Council, EU Graphene Flagship, EU FlagEra. R. Katiyar acknowledges financial support from DOE (Grant No. DEG02-ER46526) and stipend to A. Gaur. S. Sahoo acknowledges receiving financial support as PDF through NSF Grant #EPS-01002410. S. Sanvito thanks Science Foundation Ireland (grant No. 14/IA/2624) for financial support. "I.R. thanks the European Union for financial support through the FP7 project ACMOL (Grant agreement number 618082).

TOC graphic

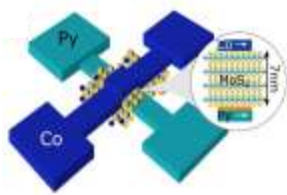 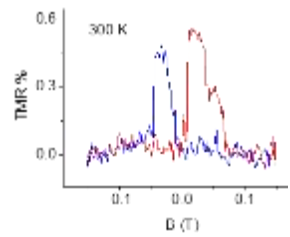